\documentclass[9pt]{article}
\usepackage{cite,subeqnarray}
\usepackage[title]{appendix}
\usepackage{epsfig,amsmath,amssymb,mathtools}
\usepackage{enumitem,setspace}
\usepackage{mathrsfs}
\usepackage{empheq}
\usepackage{soul,ulem,authblk}
\usepackage{centernot}
\usepackage[usenames,dvipsnames]{color}
\usepackage[pagebackref=true, colorlinks=true]{hyperref}
\definecolor{redish}{rgb}{0.7,0.2,0.0}  % color defined in (r=red,g=green,b=blue) model
\definecolor{bluish}{rgb}{0.2,0.5,0.8}
\hypersetup{linkcolor=redish,          % color of internal links
                  citecolor=blue,        % color of links to bibliography
                  filecolor=magenta,      % color of file links
                  urlcolor=blue}          % color of the url
\def \({\left(}
\def \){\right)}
\def \[{\left[}
\def \]{\right]}
\begin{document}
\title{The Undecidable Charge Gap and the Oil Drop Experiment}
%\author{~}
\author{Abhishek Majhi\footnote{abhishek.majhi@gmail.com}}
\affil{\small  Indian Statistical Institute,\linebreak Plot No. 203, B. T. Road, Baranagar,\linebreak Kolkata 700108, West Bengal, India.}%
%\email{abhishek.majhi@gmail.com}
%\affiliation{Indian Statistical Institute,\\Plot No. 203, Barrackpore,  Trunk Road,\\ Baranagar, Kolkata 700108, West Bengal, India\\}
\date{~}

\maketitle
%\end{document}
%\section*{Abstract}
\begin{abstract}
Decision problems in physics have been an active field of research for quite a few decades resulting in some interesting findings in recent years. However, such research investigations are based on a priori knowledge of theoretical computer science and the technical jargon of set theory. Here, I discuss a particular, but a significant, instance of how decision problems in physics can be realized without such specific prerequisites. I expose a hitherto unnoticed contradiction, that can be posed as a decision problem, concerning the oil drop experiment and thereby resolve it by refining the notion of ``existence'' in physics. This consequently leads to the undecidability of the charge spectral gap through the notion of ``undecidable charges'' which is in tandem with the completeness condition of a theory as was stated by Einstein, Podolsky and Rosen in their seminal work. Decision problems can now be realized in connection to basic physics, in general, rather than quantum physics, in particular, as per some recent claims.
 %Extrapolation of such ideas to the notion of ``quantity'' in general can potentially open up new allies concerning the foundations of metrology.   %The work is expected to impact basic physics and the foundations of metrology as far as the notion of ``quantity'' is concerned.	
 \vspace{0.2cm}
 
\end{abstract}
%\tableofcontents
%\newpage 

 \section{Introduction}   
 Considering Einstein's viewpoint\cite{einphreal}, ``{\it Physics constitutes a \ul{logical system of thought}....% which is in a state of evolution, and whose basis cannot be obtained through distillation by any inductive method from the experiences lived through, but which can only be attained by free invention. 
 	The justification (truth content) of the system rests in the proof of usefulness of the resulting theorems on the basis of \ul{sense experiences}, where the relations of the latter to the former can only be \ul{comprehended intuitively}.}'', where  
 ``sense experiences'' means ``human experience'' that occur through ``experiment and measurement''\cite{epr}.   
 Hilbert, in his sixth problem, called for {\it Mathematical Treatment of the Axioms of Physics}\cite{hilbertprob} in search of rigour and precision of reasoning in physics from the stand point of mathematical logic\cite{hilbertacker} by considering physics, like Einstein, as ``{\it a logical system of thought}''. {\color{black}{In modern days one generally terms Hilbert's call as ``{\it axiomatization of physics}'', which necessarily involves investigations concerning the ``{\it semantics of physics}''\cite{gorban} or, in simple words, the logic and language of physics\cite{ll1,lpp}. Significance of such investigations has always been realized\cite{wm1,wm2,aqft}, which has given birth to foundational problems in mathematical science\cite{massgap}.}} {\color{black} On the other hand, Einstein's recognition of the role of {\it intuition} has a parallel in the writings of Brouwer who did pronouncedly discuss the role of intuition in human reasoning in connection with the inherent undecidability associated with logic\cite{brouwer,brouwerconscious,brouwerintform}.} The word ``intuition'' means the intellect with which we construct reason, take decisions, make choices and therefore, that remains at the foundation of human reasoning and goes beyond formalism\cite{brouwerintform}. It is quite well known today, mainly due to Goedel\cite{goedelincom}(but also Turing\cite{turing1,turing2}, Church\cite{church1,church2} and others\cite{udbook}), that any logical system contains {\it undecidable} statements\cite{britannica} i.e. statements which can not be decided to be true or false within that concerned logical system -- also termed as {\it decision problem} as per Hilbert and Ackermann\cite{hilbertacker}. Naturally one can wonder whether there are undecidable statements or decision problems in physics {\color{black} and whether intuition can play a role in overcoming such obstacles of reason in relation to our sense/experimental experiences}.

  As a matter of fact, decision problems in physics have been a matter of investigation since a few decades and continues to be an active field of research {\color{black}e.g. one can consult refs.\cite{costa1,costa2,costa3,moore1,moore2,kanter1990,penrose,udqua1,udqua2,udgap1,udgap2,udgap3,udgap4,landsman2020,svozil1993,svozil1995,book2021,calude2007,calude,kaufmann2019} and the relevant references therein, which I shall refer to as the ``{\it standard literature}'' henceforth}. However, the {\it standard literature} is mostly rooted to the notion of Turing's computability and {\it halting problem} associated with his ``automatic machines'' (that excluded ``choice machines'') that was indeed an application of a decision problem, but restricted to the issue of computability of  numbers\cite{turing1,turing2}. Therefore, the research investigations of the {\it standard literature} are based upon the knowledge of computer science and the technical jargon of set theory which itself is fraught with inherent inconsistencies of reasoning\cite{fraenkel} that affect any investigation regarding the conception of truth\cite{tarskitruth,tarskiundef,tarskidef,kripketruth}.

As I plan to discuss here, decision problems in physics can be realized, without requiring any knowledge of computer science and the intricacy of set theory, in a much more elementary way by investigating the language in which experimental experience is expressed to tally with the theory. I demonstrate my viewpoint by pointing out a hitherto unnoticed contradiction, which can be posed as a decision problem,  in the oil drop experiment and thereby resolving it by {\it intuitively} refining the notion of ``existence'' (e.g. see ref.\cite{stanexistence} and the references therein), in physics in general, and of charge in particular, leading to the undecidability of the charge spectral gap.The process of reasoning is in tandem with the completeness condition of a theory as was stated by Einstein, Podolsky and Rosen (EPR)\cite{epr},{\color{black} which reveals a subtle connection between the notion of ``existence''\cite{stanexistence} and the mind-body question\cite{wigner}. % is manifested, which reveals an aspect of EPR's completeness condition discussed otherwise only in relation to quantum mechanics. Foundational issues concerning calculus and metrology (already glimpsed in ref.\cite{chupqg}) can now be directly related to EPR's completeness condition of a physical theory. 
}

 The structure of this article can be debriefed as follows. In Section(\ref{udquantum}), I focus on the relevant literature to expose the difference in approach, to deal with decision problems in physics, with the present work. Consequently, this provides the motivation for the present discussion. % to discuss, and elucidates the importance of, this work as far decision problems in physics and the role of intuition in solving such decision problems, by refining logic, is concerned. 
 In Section(\ref{oil}), I revisit the definition of electric field from a logical point of view and, in view of this, I point out how the conclusion of the oil drop experiment contradicts with its premise so as to give rise to a decision problem. In Section(\ref{res}), I discuss how such a decision problem can be resolved by refining the notion of ``existence'' in physics, which nevertheless leads to the notion of ``undecidable charges''. Also, I point out how such a resolution of the decision problem is in tandem with the completeness condition, of a theory, that was stated by EPR in their seminal article\cite{epr}. Finally, in Section(\ref{conclusion}), I conclude with some comments.

 %\section{Salient features of and recent trends concerning decision problems in physics}
 
 {\color{black}\section{Undecidability in physics -- outer inquiry vs inner inquiry/self-inquiry}  \label{udquantum}}
  
 {\color{black}The {\it standard literature} suggests that undecidability in physics can be discussed from a variety of viewpoints that depend on the use of terminologies with varying interpretations in different contexts. However, there is a difference in the underlying attitude in which the present work differs from the {\it standard literature}. I intend to explain the issues in this section.} 
 
{\color{black} \subsection{Terminologies and interpretations in the {\it standard literature}}}
 In classical physics, {\color{black} undecidability can be associated with  {\it non-computability} - explained as the algorithmic impossibility of deciding or \ul{determining} the integrability of a given Hamiltonian\cite{costa1, costa2, costa3},  {\it unpredictability} - arising in relation to the initial conditions of differential equations which, when solved (i.e. the solution is \ul{determined}), predict the trajectories of mechanical systems\cite{ moore1,moore2} (based on the methods of Lyapunov\cite{lyapunov}). Each of these two scenarios gets related to the issue of {\it determinism}, albeit applied from different point of views by the respective authors to justify ``what to determine''. Indeed such practice go beyond the categorical distinction of ``classical'' and ``quantum''\cite{popper,earman1986,earman2007,standeterminism,gisin1,gisin2,gisinhidden,santo1,santogisin1,santogisin2,yami2020}. 
In quantum physics, undecidability is associated {\it indeterminism} as well, but its essence now depends on the issues regarding entanglement, non-locality, hidden variables\cite{epr,bell,kochenspecker}, quantum measurement process\cite{udqua1,udqua2}, etc., although doubts have been cast on the accepted view of associating hidden variables uniquely to quantum physics\cite{gisinhidden}. 
Investigations regarding {\it determinism} (or {\it indeterminism/indeterminacy}) in quantum physics, one hand, have consequences concerning quantum many body systems\cite{udgap1,udgap2,udgap3,udgap4} and, on the other hand, lead to issues related to free will and choice\cite{landsman2014,kochen2017,heyred1983,conkoch2006,conkoch2009}.  To mention, {\it indeterminism} is often identified to {\it randomness}\cite{landsmanrandom,svozil1993}. 

Such observations regarding the {\it standard literature} is just suggestive of the fact that how such terms are used by any concerned researcher in any particular context, and associated with undecidability in physics, is itself open to discussion and debate and appears to be far from settled. Of particular interest in the present context is the fact that, ignoring such room for doubt, in recent past it has been claimed that decision problems are unique features of  quantum physics} as I point out in what follows.  
\begin{itemize}
\item Refs.\cite{udqua1,udqua2} convey the message that undecidable problems have some particular connection to quantum physics only. For example, the authors of ref.\cite{udqua2} write in the abstract itself that ``{\it apparently simple problems in quantum measurement theory can be undecidable even if their classical analogues are decidable. Undecidability hence appears as a genuine quantum property here.}''
\item Refs.\cite{udgap1,udgap2,udgap3,udgap4}, epitomized by the mass gap problem of ref.\cite{massgap}, convey the message that the spectral gap problem is uniquely associated with quantum theory. Thus, the fact that it is an undecidable problem, which is the central point of emphasis of refs.\cite{udgap1,udgap2,udgap3,udgap4}, attains its validity only in association with quantum physics.
\end{itemize}
Such philosophy of connecting decision problems uniquely to quantum mechanics has its roots in a popular book of Penrose\cite{penrose} which has nevertheless been counteracted to certain extent in ref.\cite{costa2}. {\color{black} In a nutshell, whether there is a unique notion of `undecidability' that can be attached to physics, seems to be indecisive from a study of the {\it standard literature}.}

{\color{black}\subsection{{\it Standard literature} and outer inquiry}

Nevertheless, one can decisively assert that the {\it standard literature} is an ongoing amalgamation of physics (written in informal language) and mathematical logic/theoretical computer science (deals with formalization of language), to study which one} requires a fairly detailed knowledge of set theory and mathematical logic, relating to the works of Goedel\cite{goedelincom} and Turing\cite{turing1,turing2} concerning the foundation of theoretical computer science\cite{cs1,cs2}. {\color{black} The core essence of theoretical computer science is to mechanize or arithmetize reasoning and, therefore, to compute with the language. Thus, it is all about the structure of the language and the respective algorithm for computation. 
%Therefore, in the {\it standard literature} the discussion  concerning undecidability in physics is basically the realization of the same by feeding the statements of physics into algorithmic structure of computer programming language. The attitude is well manifested from the following phrase:  ``{\it the Halting Problem for Turing Machines can be encoded in the spectral gap problem}''\cite{udgap2}. 
This is just an implementation of Hilbert's philosophy underlying his sixth problem\cite{hilbertprob}, the essence of which becomes evident from the following words of Hilbert and Ackermann, on page no. 5 of ref.\cite{hilbertacker}: ``{\it .... the truth or falsehood of a sentential combination depends solely upon the truth or falsehood of the sentences entering into the combination, and \ul{not upon their content}.}'' So, the {\it standard literature} investigate {\it what we can do with the statements of physics by mechanizing them in the form of a computer programming}, without worrying about the inner content of those statements. This is what I may call {\it outer inquiry}, as one is {\it ignorant} about the inner structure of the statements of physics which are always subject to refinement as far as the correspondence with sense experience is concerned
(e.g. see ref.\cite{ll1} for a flavour of what it means to analyze the inner content) and, consequently, the subtleties of reasoning associated with the foundations of quantity calculus\cite{mills1997,boer1995,williams2014,emerson2004,guggenheim1942,mari2009,marietal2012,mills1997,wolff,ehrlich2007}, which deals with the language of physics\cite{vim2007,vim2021}, are not considered to be of interest. %Essentially, this is an investigation regarding {\it what we can do with the statements of physics by mechanizing them in the form of a computer programming.} This is what I may call {\it outer inquiry}, as we are ignorant about the inner structure of those statements which are always subject to refinement as far as its correspondence with sense experience is concerned. So, the terms like ``quantity'', ``differential equation'', etc. are considered as perfectly understood and accepted without question, to form the premises of the {\it standard literature}. 

While such {\it ignorance} may not be considered as a drawback of the {\it standard literature}, but certainly it reveals a legitimate unexplored route of inquiry into the foundations of physics that concerns the inner architecture of the statements in correspondence with sense experience\cite{ll1,lpp}. This is what I discuss next. }

%	This is equivalent to the ignorance of any foundational question that may potentially arise as far as the correspondence between human sense experience and the language of physics is concerned e.g. see refs.\cite{ll1,lpp}.
% that has led to interesting realizations -- such as the direct link of the Heisenberg uncertainty principle to Cauchy's definition of derivative\cite{lpp,chupqg}, formal unprovability of the first Maxwell's equation in light of EPR's completeness condition of a physical theory\cite{maxwellepr} --  and potential solutions to serious problems of physics, such as a non-singular theory of two body gravitational interaction through a refinement of the first axiom of geometry along with a demonstration of the undecidability of the continuum hypothesis\cite{dot} and a glimpse of a possible non-singular theory of interaction between two charged bodies\cite{maxwellepr}.

{\color{black} \subsection{Language of physics and inner/self-inquiry}
Language is the only mean of our expression and communication regarding ``truth''\cite{tarskitruth,tarskidef,tarskiundef,kripketruth}, which is directly based upon our sense experiences and experimental observations as far as physics is concerned\cite{einphreal}. Therefore, refinement of language and logic, in connection to operation becomes critically important in the pursuit of such truth\cite{ll1,lpp}. This is evident, as well, from Bohr's writings on p 3 of ref.\cite{bohr1958}:  ``{\it ... the description of the experimental arrangement and the recording of observations must be given in \ul{plain language, suitably refined by the 	usual physical terminology}. This is a simple logical demand, since by the word ``experiment'' we can only mean a procedure regarding which we are able to communicate to others what we have done and what we have learnt.}''. 

Bohr's ``plain language'', which can be likened to what Tarski called ``natural language''\cite{tarskitruth}, is the ``free invention'' in Einstein's words (quoted earlier). %Further, I may point out that what we mean by ``physical dimension'' in physics (viz. ``mass''$[M]$, ``length''$[L]$, ``time''$[T]$, etc.)\cite{vim2007,vim2021}, represent ``physical terminology'' in Bohr's words (quoted above) and ``physical lexicons'' in Quine's words\cite{quinelexicon}. 
What Bohr demands as a suitable refinement, is actually an intuitive process of establishing the linguistic expression that represents the sense experience as closely as possible. The significance of such demand has been elaborately discussed in refs.\cite{ll1,lpp} that can potentially have a radical impact on the foundations of physics viz. a direct realization of uncertainty principle from the definability of ``derivative''\cite{chupqg}, unprovability of the first Maxwell's equation in light of EPR's completeness condition\cite{maxwellepr}, a non-singular structure of two body gravitational interaction with a demonstration of the undecidability of the continuum hypothesis\cite{dot}. 

I may call such a line of investigation as {\it inner inquiry} or {\it self-inquiry} as it is concerned with 
\begin{enumerate}
\item {\it how truthfully the language of physics corresponds to our sense experience and experimental observation (formation of natural/plain language through free invention),}
\item {\it how much of the natural/plain language can be put in a computable form (extraction of the computational content).}	
\end{enumerate}
One may consult refs.\cite{ll1,lpp,chupqg,maxwellepr,dot} how such a line of inquiry radically affects the foundations of physics. To mention, I have borrowed the term ``inner inquiry'' from Brouwer  on p 494 of ref.\cite{brouwer} (or ref.\cite{brouwerconscious}).%: ``{\it ....intuitionism on the one hand subtilizes logic, on the other hand denounces logic as a source of truth. Further that intuitionistic mathematics is inner architecture, and that research in foundations of mathematics is \ul{inner inquiry} with revealing and liberating consequences, also in non-mathematical domains of thought.}''
%Here, I may provide two examples to demonstrate what I mean by inner/self-inquiry with the language of physics -- the first one reveals the difference in attitude regarding what questions to ask and the second one is an explicit demonstration concerning. 

}

%{\color{black}\subsubsection{The difference in attitude}	
	
%``{\it We must also be precise in what we mean by ``gapped'' and ``gapless''... Since quantum phase transitions occur in the thermodynamic limit of arbitrarily large system size, we are interested in the spectral gap 	$\Delta(H_L)=\lambda_1(H_L)-\lambda_0(H_L)$ as the system size $L\to\infty$ (where $\lambda_0,\lambda_1$ are the lowest and second-lowest eigenvalues). We take ``gapped'' to mean the system has a unique ground state and a constant lower bound on the spectral gap: $\Delta(H_L)\geq \gamma >0$ for all sufficiently large $L$. We take ``gapless'' to mean the system has \ul{ continuous} spectrum above the ground state in the thermodynamic limit.}''
	
%}

{\color{black}\subsection{Inner/self-inquiry and decision problem (``problem of universal validity'')}}

{\color{black} Decision problem is a part of human reasoning in general\cite{savage,decision1,decision2} and it was aptly emphasized by Brouwer through the following statements on p 111 of ref.\cite{brouwer}: ``{\it In wisdom there is no logic. In science logic often leads to the right result, but it cannot be trusted to do so if its application is indefinitely repeated. In mathematics it is uncertain whether the whole of logic is admissible and it is uncertain whether the problem of its admissibility is \ul{decidable}.}'' As a matter of fact, Turing's works\cite{turing1,turing2} that founds the basis of the {\it standard literature}, is about how to practically realize Hilbert's decision problem, through a working machine (automatic machine, not a choice machine) , as discussed in Chapter III of the book by Hilbert and Ackermann\cite{hilbertacker}.} 
So, following Hilbert and Ackermann, from p.112 of ref.\cite{hilbertacker}, I may call a ``{\it decision problem}'' as a ``{\it problem of the universal validity}'' of some formal statement i.e. I investigate  whether we can keep a formal statement to be universally valid throughout a process of reasoning while constructing a theory of physics and drawing experimental conclusions from that. Therefore, if I consider the notion of a ``decision problem'' associated with Turing's computability as algorithmic and artificial in nature, then my consideration of the notion of ``decision problem'' is founded on intuitive refinement of the epistemological rigour of the axioms themselves which are stated by us (humans). While the former is outer inquiry that aims to fit the input of a machine like that of Turing, the latter is inner (self-)inquiry that is performed by a human so as to choose and refine the axioms, in connection to sense experience, through the execution of intellect and intuition. %{\color{black} Let me provide some explicit examples to explain how self-inquiry differs from outer inquiry of the {\it standard literature} before I deal with the decision problem to be discussed in the present work.}

% If the {\it standard literature} concerns what is computable or not in physics from the computer science point of view, then the present discussion concerns how truthfully does the language (i.e.  the axioms, postulates, definitions, etc.) of physics express our sense experience and how much of it is computable. 

{\color{black}In Appendix (\ref{appA}), I have provided an example of the second aspect of self-inquiry i.e. extraction of the computational content of the statements of physics.} Here, I intend to discuss that the spectral gap problem is undecidable in its own right, {\color{black}which will exemplify the first aspect of self-inquiry i.e. how truthfully the statements of physics correspond to our sense experience.} Neither the spectral gap problem nor the problem of undecidable statements in physics has a priori any unique connection to quantum physics; rather decision problems are part of human reasoning and therefore, plague science as a whole e.g. see refs.\cite{savage, decision1, decision2}, apart from ref.\cite{hilbertacker}, for the general scenario and see refs.\cite{amrr1,amrr2,buddhamath} for particular elementary examples. {\color{black} It is only that there can be different types of realization of undecidability in physics, of which, the present discussion is about a type that differs from the rest of the literature right from the outset.} I expose my viewpoints through an analysis of the principles that underlie the oil drop experiment\cite{millikan,fletcher}, which is expected to impact the reasoning of basic physics that is a concern of everyday practice of the physicist.

 \section{The oil drop experiment and a decision problem}\label{oil}
 To elucidate my concern, I consider the example of the famous oil drop experiment that, was performed by Millikan\cite{millikan} and Fletcher\cite{fletcher} and, can be considered as one of the most important experimental epitomes of modern physics. I discuss how the interpretation of the oil drop experiment is founded on a hitherto unnoticed contradiction, which can nevertheless be posed as a decision problem.

\subsection{Definition of electric field}
I begin by revisiting the standard definition of electric field so as to highlight the constituent logical elements. The definition of electric field is given by considering the Coulomb's law as a premise that is stated as follows. The magnitude of the force $(F)$ between two objects having charges $(q_1, q_2)$, which are at rest with respect to each other, is proportional to the product of the two charges and inversely proportional to the square of the distance between those two point charges. This statement is expressed as 
\begin{eqnarray}
	F=k \frac{q_sq_t}{r^2}\label{claw}
\end{eqnarray}
where $k$ is a proportionality constant that needs to be fixed by experiment and $r$ is the distance between the two charges, each considered as a point. Now, the electric field due to $q_s$ (source charge), is {\it defined} as
\begin{eqnarray}
	E(q_s):=\lim_{q_t\to 0}\frac{F}{q_t}\quad: F=k \frac{q_sq_t}{r^2}.\label{q1to0}
\end{eqnarray}
$q_t$ is called the test charge (e.g. see page no. 16 of ref. \cite{purcell}, page no. 29 of ref.\cite{schwartz}, page no. 25 of ref.\cite{rmc}).  The symbolic statement ``$\lim_{q_t\to 0}$'' conveys the meaning that ``{\it the test charge $q_t$ is chosen to be arbitrarily small}''. This requirement of the limiting condition is necessary for the definition of an electric field due to any arbitrary charge configuration or distribution as well (however, see ref.\cite{ll1} for an exposition of the associated subtleties in reasoning). Millikan used the concept of electric field explicitly in ref.\cite{millikan} -- he called it ``electric field strength'' and denoted it by the symbol ``$F$'', rather than the symbol ``$E$'' that I use here according to the modern convention.

 \subsection{A contradiction or a decision problem}
 Now, having revisited the definition of electric field, I consider the following statement whose logical truth is to be investigated.
\begin{center} {\bf M:} {\it Arbitrarily small charge does, hence a charge gap does not, exist.} 
\end{center} 
 The decision problem that concerns the oil drop experiment is posed through the following question:
\begin{center}
{\bf Q:} {\it  Can it be {\it decided} whether {\bf M} is true or false, considering the interpretation of the oil drop experiment to be a logical truth?}
\end{center}
To explain that the answer to {\bf Q} is in the negative, I point out the following two possibilities, of which the first one concerns the conclusion and the second one concerns the premise of the oil drop experiment.
\begin{itemize}
\item {\bf M} {\it is false. It is a conclusion of the oil drop experiment that a smallest charge, hence a charge gap, exists.~~~~} 
\item {\bf M} {\it is true. Electric field can be defined if and only if arbitrarily small charge does, hence a charge gap does not, exist. } 
\end{itemize}
Since the theoretical analysis of the oil drop experiment in ref.\cite{millikan} is based on the use of the concept of ``electric field'' (i.e. truth of {\bf M}) and the respective interpretation of the experimental observations lead to the conclusion of the existence of a smallest (``elementary'') charge (i.e. falsehood of {\bf M}), hence, the whole process of reasoning associated with the oil drop experiment does not let us decide whether a charge gap exists or not because {\bf M} is both true and false in the same process of reasoning that underlies the interpretation of the oil drop experiment. Therefore, a hitherto unnoticed contradiction underlies the interpretation of the oil drop experiment. The situation can be viewed in an alternative way. That is, the following scenario arises if one decides to consider any one of the above two answers to be universally valid throughout the whole process of reasoning. 
\begin{itemize}
\item {\bf Decision 1 (D1):} {\bf M} is false. Therefore, the conclusion of the oil drop experiment is logically true, but  the premise i.e. the definition of electric field, that leads to  such a conclusion, is logically false.
\item  {\bf Decision 2 (D2):} {\bf M} is true. The conclusion of the oil drop experiment is logically false.
\end{itemize} 
{\bf D1} renders the notion of ``electric field'' to be undefined and, hence, the symbol ``$E$'' meaningless.  {\bf D2} jeopardizes the logical validity of the conclusion of the oil drop experiment. Thus, in totality, the logic underlying the interpretation of the oil drop experiment is plagued by a decision problem and this explains why the answer to {\bf Q} is in the negative. 

\section{Resolution of the contradiction or the decision problem}\label{res}
   Now, the matter of investigation is to find a way to resolve the contradiction or, equivalently, to find a solution to the decision problem. I propose a possible way to do so by refining the statement {\bf M} itself.  I begin by dismissing the validity of the question {\bf Q} by claiming the {\it incompleteness} of {\bf M}. To be more specific, the notion of ``{\it existence}'' is not completely specified in {\bf M}. Therefore, {\bf M} {\it is neither true nor false}. This renders {\bf Q} to be a logically invalid question and then there is no decision to be made i.e. no decision problem. So, the necessary task is to {\it complete} the statement {\bf M} by specifying the {\it type} of {\it existence}. To do that, at first, I need to consider the following postulate.\vspace{0.2cm}

{\bf Postulate on Existence (P$_{EX}$):} {\it There are two types of ``existence'' in physics. One is ``theoretical existence'' which is what we assume to exist in terms of language and expressions to write a theory. The other one is ``experimental existence'' which  is what we can experimentally verify considering the theory to be a mean of interpretation of the experiment.}\vspace{0.2cm}

From {\bf P$_{EX}$} it is now evident how {\bf M} is a logically incomplete statement and that {\bf M} can be completed in two ways, leading to two distinct statements as follows.
\begin{itemize}
\item {\bf M$_{T}$:} {\it Arbitrarily small charge does, hence a charge gap does not, exist theoretically.}
\item {\bf M$_E$:} {\it Arbitrarily small charge does, hence a charge gap does not, exist experimentally.}
\end{itemize}
Thus, {\bf M} loses any meaning in itself and it is now refined into two different statements {\bf M$_T$} and {\bf M$_E$}. %However, added to this, I need to consider a further postulate as follows.\vspace{0.2cm}
%{\bf Postulate 2 (P2):} {\it Considering a quantity of charge, written in terms of symbols to formulate a theory of physics, for any charge denoted by $q$, theoretically there always exists some charge $q'$ such that $q'<q$, or equivalently, $q'=\epsilon q~: 0<\epsilon <1$.}\vspace{0.2cm}

Now, by {\bf P$_{EX}$} and {\bf M$_{T}$} the test charge for the definition of electric field can be chosen arbitrarily small in theory. When the experiment yields the smallest detectable charge, say $e$, then {\bf M$_{E}$} is falsified. Thus, in the whole process of reasoning, that underlies the interpretation of the oil drop experiment, {\bf M$_{T}$} is true and {\bf M$_{E}$} is false. There is neither any contradiction nor any decision problem.

\subsection{Undecidable Charges and the EPR Completeness Condition}
Now, I may note that theoretically there exist charges ($Q_u$-s), by {\bf M$_{T}$}, such that for any $Q_u$ the following condition holds: $Q_u<e$, or equivalently,  $Q_u=\epsilon_u e~: 0<\epsilon_u<1$. 

%I call such $Q_u$-s as {\it undecidable} charges -- both in theory and in experiment. 

Since $e$ represents the charge such that any charge smaller than $e$ can not be experimentally detectable and hence, the experimental existence of $Q_u$-s can neither be proven true nor be proven false i.e. existence of $Q_u$-s is experimentally undecidable. 

Now, there remains the question of theoretical undecidability i.e. considering {\bf M$_{T}$}, or theoretical existence of $Q_u$-s, as a  ``hypothesis'' whether it can be proven to be true or false within the theory. This question can be answered as follows. Using {\bf M$_{T}$},  $Q_u$-s are ``hypothesized'' to define electric field, founded on which is the theoretical construction. Therefore, the theoretical construction is built upon the ``hypothesis'' of $Q_u$-s i.e. {\bf M$_{T}$}. Now, if one wonders whether there is a theoretical proof or disproof of this ``hypothesis'', the answer is obviously negative in either cases. This is because the ``hypothesis'' is the only reason that let us write down the theory itself by avoiding any logical fallacy. Any attempt either to prove or to disprove the ``hypothesis'' theoretically is like using the premise to either prove or disprove itself. A successful proof of the ``hypothesis'' leads to a useless tautology: 
\begin{eqnarray}
	\text{``{\it $Q_u$ exists theoretically, if the theoretical existence of $Q_u$ is hypothesized to construct the theory}.''}\nonumber
\end{eqnarray} 
And, a successful disproof of the same leads to a useless contradiction:
\begin{eqnarray}
\text{``{\it $Q_u$ can not exist theoretically,  if the theoretical existence of $Q_u$ is hypothesized to construct the theory.}''}\nonumber
\end{eqnarray} 
Therefore, the ``hypothesis'' is theoretically neither provable nor unprovable i.e. undecidable. This is why the theoretical existence of the undecidable charges ($Q_u$-s) is a undecidable hypothesis i.e. a postulate.  It is only such a postulate that lets us logically define the electric field and construct the theory which in turn provides the underlying framework to analyze the experimental data of the oil drop experiment that leads to the experimental existence of $e$.

Now, let me explain how {\bf M$_{T}$} and {\bf M$_{E}$}, and hence the postulate of $Q_u$-s (undecidable charges) is actually a different way of stating the completeness condition of a {physical}%\footnote{\sout{Unlike EPR, I remain more conservative by not judging, a priori, any theory to be physical or unphysical and let it remain open to be decided by the experimental verification. Therefore, in contrast to EPR, I drop the adjective ``physical'' and consider the word ``theory'' instead of ``physical theory''.}} 
theory as was written by EPR\cite{epr}, which I may call EPR's Completeness Condition (ECC):\vspace{0.1cm}

 ``{\it every element of the physical reality must have a counterpart in the physical theory}''. \vspace{0.1cm}

%For that, I may consider the following excerpt from ref.\cite{epr}.``{\it In attempting to judge the success of a physical theory, we may ask ourselves two questions: (1)``Is the theory correct?'' and (2) ``Is the description given by the theory complete?'' .... %It is only in the case in which positive answers may be given to both of these questions, that the concepts of the theory may be said to be satisfactory. 	The correctness of the theory is judged by the degree of agreement between the conclusions of the theory and human experience. This experience, which alone enables us to make inferences about the reality, in physics takes the form of experiment and measurement. .... %It is the second question that we wish to consider here, as applied to quantum mechanics.  Whatever the meaning assigned to the term complete, the following requirement for a complete theory seems to be necessary one: every element of the physical reality must have a counterpart in the physical theory. We shall call this condition of completeness. The second question is thus easily answered, as soon as we are able to decide what are the elements of the physical reality.}''

%{\small{\it The elements of the physical reality cannot be determined by a priori philosophical considerations, but must be found by an appeal to results of the experiments and measurements. A comprehensive definition of reality is, however, unnecessary for our purpose.....}	''

It is quite obvious that ECC is a logical implication rather than a logical equivalence i.e. {\it every element of the theory may NOT have a counterpart in the physical reality.} This can be realized as the fact that the undecidable charges are not parts of the physical reality that can be realized through experiment and measurement. However, these undecidable charges are required to define the concept of ``electric field'' and hence, to construct the theory that underlies the explanation of the experimental data of the oil drop experiment leading to the conclusion about what physical reality is. 

{\color{black} Interestingly, such a process of reasoning reveals a clear and direct link between ECC and the mind-body distinction\cite{wigner,ll1}, in the pursuit of truth and reality, as follows. The postulate of undecidable charges belongs to the theory that is constructed to interpret experimental observation realized through measuring instruments and sense experience. The theory being a construction of our mind represents the mind-aspect. Experimental observations being performed through sense experiences represents the body-aspect. Thus, it is the human mind that constructs the undecidable which lays the foundation to test and decide the physical reality perceived through the sense experiences of the human body. Thus, the undecidable charge(s) is the ``free invention'' (in Einstein's words, quoted earlier) that makes the reasoning of the oil drop experiment consistent. In passing I may note that the present course of reasoning, based on self-inquiry, is much more elementary than the discussions connecting ``free will''\cite{conkoch2006,conkoch2009,landsman2014} and ``hidden variables''\cite{bell} rooted to the EPR paper\cite{epr}, which are based on outer inquiry. I shall keep further elaboration on this matter for another occasion as it may result in too much deviation from, thus affecting the simplicity of,  the main issue of discussion, due to a possible association with psychology\cite{piaget}, psychophysics\cite{psychophysics} and psychoanalysis\cite{machanalysis}, which may in turn involve a discussion concerning human psychophysical perception of time\cite{psychtime} and its possible role in shaping our mathematical language\cite{gisinnature}.}

%I have used the word ``postulate'' rather than the word ``hypothesis'' i.e. a postulate is a premise on which a theory is built and a hypothesis is a statement that is left open to be verified. % because it is independent of the axioms of the concerned theory. %{\bf M$_{T}$} 
%\section{Turing's ``$a$-machine'' and ``$c$-machine''}
%``{\it For some purposes we might use machines (choice machines or $c$-machine) whose motion is only partially determined by the configuration (hence the use of the word "possible" in §1). When such a machine reaches one of these ambiguous configurations, it cannot go on until some arbitrary choice has been made by an external operator. This would be the case if we were using machines to deal with axiomatic systems. In this paper I deal only with automatic machines, and will therefore often omit the prefix $a$-.}''

\section{Conclusion}\label{conclusion}
Decision problems are not only part of mathematical logic\cite{hilbertacker}, but also part of human reasoning in general\cite{savage,decision1,decision2}. While the mathematical treatment of the axioms of physics is required for clarity of reasoning\cite{hilbertprob}, intuitive refinement of the elements of logic is also required for precision of expressions that directly correspond to human ``sense experiences''\cite{einphreal} -- a process that may be termed as ``inner inquiry''\cite{brouwerconscious} or self-inquiry\cite{ll1,lpp,chupqg}. The resolution of the hitherto unnoticed decision problem, which plagues the underlying explanation of the oil drop experiment, showcases one such example where a decision problem is resolved through intuitive refinement of a postulate regarding the existence of quantity by categorizing the notion of ``existence'' into two {\it types}. Also, such refinement is in tandem with the completeness condition of a theory that was stated by Einstein, Podolsky and Rosen\cite{epr}. In view of this it becomes strongly manifest that decision problems in physics and the undecidability of the spectral gap have a priori no connection to quantum mechanics unlike what has been the underlying message of refs.\cite{udqua1,udqua2,udgap1,udgap2,udgap3} (although the other aspects of such literature are novel in their own rights). While decision problems in general physics have also been discussed earlier in ref.\cite{costa1,costa2,costa3,moore1,moore2}, alongside the more recent refs.\cite{udqua1,udqua2,udgap1,udgap2,udgap3}, but the present work differs by its demonstration of the fact that decision problems in physics can be understood without the prerequisites of theoretical computer science and the technical jargon of set theory (which itself is fraught with inherent logical inconsistencies\cite{fraenkel}). While this work showcases a decision problem that directly relates to basic physics, on the other hand, the refinement of the notion of ``existence'' in physics now can potentially open the door for new questions concerning the definition of ``quantity'' as far as the foundations of physics and, in particular, metrology is concerned\cite{bipm}. I plan to discuss such issues in near future.

  {\it Acknowledgment:} The author has been supported by the Department of Science and Technology of India through the INSPIRE Faculty Fellowship, Grant no.- IFA18- PH208.\vspace{0.1cm}
 
  {\it Conflict of interest statement:} The author declares that there is no conflict of interest. \vspace{0.1cm}
 
  %{\it Declaration on competing interest:} The authors declare that there is no competing interest.\vspace{0.1cm}
 
 %{\it Data availability statement:} Data sharing not applicable to this article as no datasets were generated or analysed during the current study.\vspace{0.1cm}
 
\appendix
{\color{black}\section{Exemplifying ``problem of universal validity'' with the continuity equation}\label{appA}	
	Here I provide an example of how one can realize decision problem as ``problem of universal validity'' in physics through self-inquiry (also see the Appendix B of  ref.\cite{maxwellepr}) by studying the proof of the continuity equation that appears in standard textbooks and, otherwise, accepted beyond doubt from a logical point of view. This is actually an example of the second aspect of self-inquiry i.e. extraction of the computational content.

	The continuity equation, the so called local conservation law, forms the basis of our general understanding of flow of quantities like charge, mass, etc. in physics e.g. see refs.\cite{batchelor,marsden,landau,griffiths,jackson}. Here, I point out how the continuity equation is founded on a decision problem that allows for making suitable choices concerning two apparently contradictory propositions.  For the present discussion it is convenient to consider the continuity equation involving charge and current\cite{griffiths}. To understand the issue at hand, it is sufficient and convenient to discuss the flow of charge along a wire that can be written as follows:
	\begin{eqnarray}
		\vec
		\nabla\cdot\vec J+\frac{\partial \lambda}{\partial t}=0,\label{con}
	\end{eqnarray}
	where $\vec J=\lambda\vec v$, $\vec v=d\vec \ell/dt$, $d\vec{\ell}={dx}~\hat i+{dy}~\hat j+{dz}~\hat k$ and $\vec v$ satisfies the condition $\vec{\nabla}\cdot\vec v=0$\cite{landau, batchelor, marsden}; $\lambda[x(t), y(t), z(t), t]$ is the linear charge density (charge per unit length) at some instant $t$, $\vec J[x(t), y(t), z(t), t]$ is the line current density and $[x(t), y(t), z(t)]$ denote the spatial coordinate of any point on the line of flow at some time $t$. The construction of the relation ``$\vec J=\lambda \vec v$'' is based on two scenarios of interpreting the charge flow that can be demonstrated through figure (\ref{fig}), which can be generalized for surface and volume flows e.g. see ref.\cite{griffiths}. 
	\begin{figure}[hbt]\label{fig}
		\begin{center}
			\includegraphics[scale=0.40]{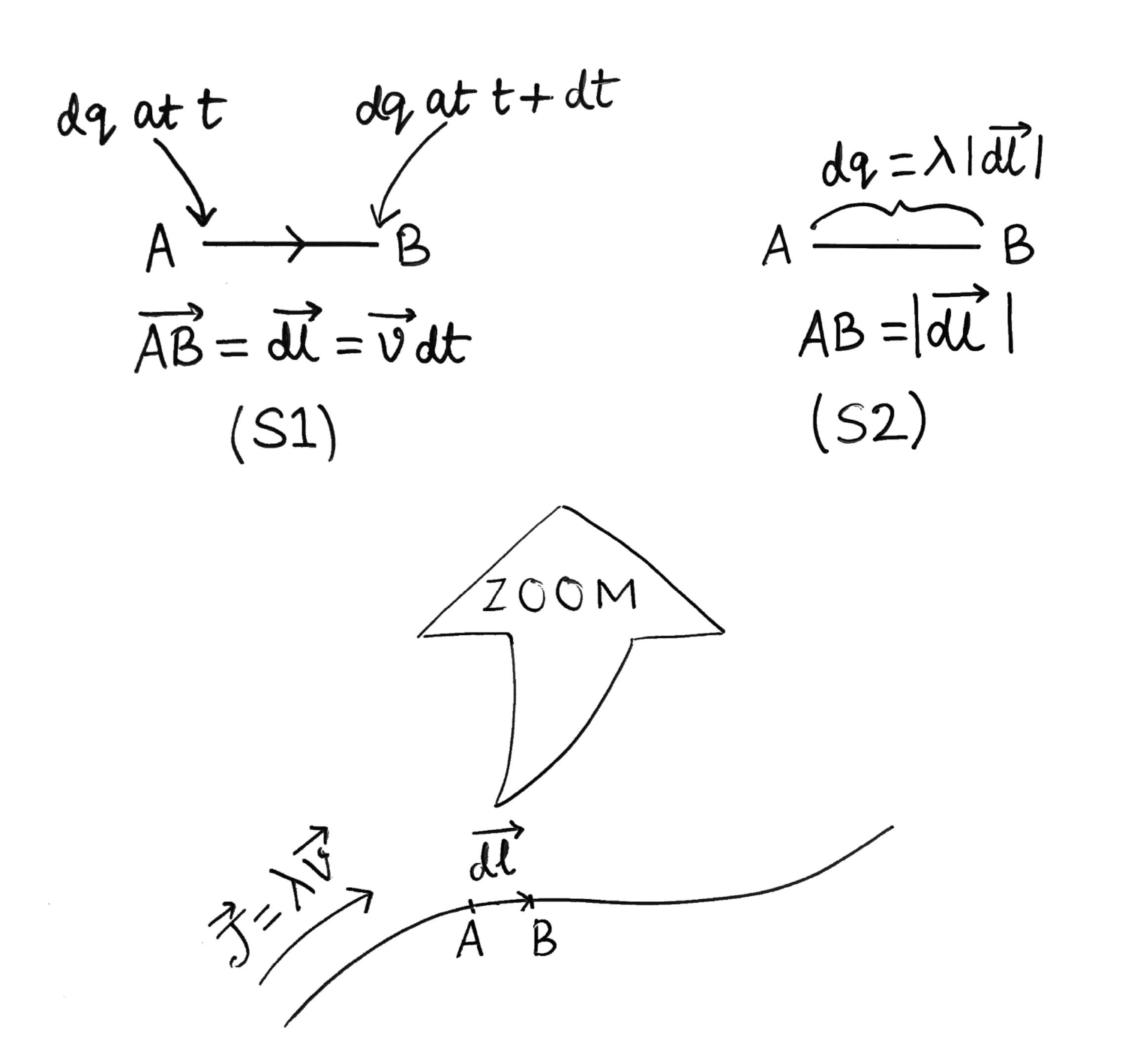}
		\end{center}
	\end{figure}
	\begin{itemize}
		\item {\bf Scenario 1 (S1):} Considering that a point charge $dq$ is displaced through $\vec{dl}$  in time $dt$ such that we can write  $\vec{dl}=\vec{v}dt$.  Then, we can write $|\vec{dl}|=|\vec{v}|dt$ and hence, $dq=\lambda |\vec{v}|dt$. In this case, the charge $dq$ is at the point $A$ at some time $t$ and then it is at the point $B$ at some time $t+dt$, so that the displacement in time duration $dt$ is $\vec{AB}=\vec{dl}$. In such a scenario the following assertion holds true:
		\begin{eqnarray}
			M \equiv \text{$dq$ is a point charge}.
		\end{eqnarray}
		
		\item {\bf Scenario 2 (S2):} At any instant $t$ of the passage of a line current, there is a line charge density $\lambda$ (charge per unit length) such that the charge $dq$ distributed over the length $d\ell$ is written as $dq=\lambda dx$. In this case, the charge $dq$ is neither at point $A$, nor at point $B$, but it is spread over the line $AB$. Therefore, the following proposition holds true for such explanation:
		\begin{eqnarray}
			\neg M \equiv \text{$dq$ is NOT a point charge}.
		\end{eqnarray}
	\end{itemize}
	Therefore, the relation $\vec{J}= dq/dt=\lambda\vec{v}$ holds true if and only if $M$ and $\neg M$ in one and the same process of reasoning that leads to such a construction. It may be written as
	\begin{eqnarray}
		\vec{J}=\lambda\vec{v}~: M\wedge\neg M. \label{conlogic}
	\end{eqnarray}
    Now, I may note that, considering $\neg\neg M\equiv M$ (classical negation), $(M\wedge\neg M)\equiv\neg(\neg(M\wedge\neg M))$, meaning the violation of the law of non-contradiction. Further, I note that  $M\wedge\neg M\equiv\neg(\neg M\vee M)$, meaning the violation of the law of excluded middle. That is, (\ref{conlogic}) can be recast as
    \begin{eqnarray}
    	\vec{J}=\lambda\vec{v}~: \neg(M\vee\neg M). 
    \end{eqnarray}
 Hence, the relation $\vec{J}= dq/dt=\lambda\vec{v}$ holds true if and only if neither $M$ not $\neg M$ (i.e. $dq$ is neither a point charge nor not a point charge) in one and the same process of reasoning that leads to such a construction.

	Therefore, the situation can be viewed as a decision problem where it can not be decided whether $dq$ is a point charge or not i.e. this is a ``problem of universal validity'' of the formal statement $M$\cite{hilbertacker}. This, however, allows for suitable choices to be made by humans (us) according to context i.e. $dq$ is considered as a point charge in {\bf S1} and it is not considered as a point charge in {\bf S2}. To mention, such a choice can not be made by a Turing machine which is, by definition, an automatic machine that excludes the scope any choice\cite{turing1}.

Interestingly, this brings forward further questions regarding the notion of ``infinitesimal quantity''. Dedekind would go as far as writing, on page no. 1 of ref. \cite{dedekindnum}, that the geometrically founded  ``introduction into the differential calculus can make no claim to being scientific'' and would decide to keep searching for an ``arithmetic'' foundation of the same. The proof of the continuity equation, as discussed above, justifies Dedekind's criticism and reveals that the notion of ``infinitesimal quantity'' gives rise to undecidability when axiomatized in terms of geometry (axiom of point). So, naturally the question arises --  Do we need new postulates or axioms to explain ``infinitesimal quantity''? Do we need an arithmetic foundation of differential calculus? What are the meanings of the words ``infinitesimal'' and ``quantity'' in connection to operations? Does the definition of ``quantity'' provided in quantity calculus be of any help regarding such issues? 

Certainly few of these questions have been addressed in ref.\cite{ll1,lpp,chupqg} and have indirect connection to ref.\cite{dot}. The present discussion regarding the oil drop experiment only showcases how subtle is the question regarding ``quantity'' (its existence).
}

%The structure of this article can be debriefed as follows. In Section(\ref{oil}), I revisit the definition of electric field from a logical point of view and, in view of this, I point out how the conclusion of the oil drop experiment contradicts with its premise so as to give rise to a decision problem. In Section(\ref{res}), I discuss how such a decision problem can be resolved by refining the notion of ``existence'' in physics, which nevertheless leads to the notion of ``undecidable charges''. Also, I point out how such a resolution of the decision problem is in tandem with the completeness condition, of a theory, that was stated by Einstein, Podolsky and Rosen (EPR) in their seminal article\cite{epr}. Finally, in Section(\ref{conclusion}), I conclude with some necessary comments. 


\begin{thebibliography}{777}
\bibitem{einphreal}A. Einstein, {\it Physics and Reality}, Journal of the Franklin Institute,
Volume 221, Issue 3, Pages 349-382 (1936), \href{https://www.sciencedirect.com/science/article/abs/pii/S0016003236910475}{https://www.sciencedirect.com/science/article/abs/pii/S0016003236910475}.


\bibitem{epr} A. Einstein, B. Podolsky, N. Rosen, {\it Can Quantum-Mechanical Description of Physical Reality Be Considered Complete?}, Phys. Rev. 47, 777 (1935),    \href{https://journals.aps.org/pr/abstract/10.1103/PhysRev.47.777}{https://journals.aps.org/pr/abstract/10.1103/PhysRev.47.777}.

\bibitem{hilbertprob} D. Hilbert, {\it Mathematical Problems}, Bull. Amer. Math. Soc. 8(10): 437-479 (1902); \href{https://projecteuclid.org/journals/bulletin-of-the-american-mathematical-society-new-series/volume-8/issue-10/Mathematical-problems/bams/1183417035.full}{https://projecteuclid.org/journals/bulletin-of-the-american-mathematical-society-new-series/volume-8/issue-10/Mathematical-problems/bams/1183417035.full}	
	
\bibitem{hilbertacker} D. Hilbert, W. Ackermann, {\it Principles of Mathematical Logic}, Chelsea Publishing Company (1950).	

\bibitem{gorban}A. N. Gorban, {\it Hilbert’s sixth problem: the endless road to rigour}, Phil. Trans. R. Soc. A volume 376, issue 2118, 20170238 (2018), \href{https://doi.org/10.1098/rsta.2017.0238}{https://doi.org/10.1098/rsta.2017.0238}, \href{https://arxiv.org/abs/1803.03599}{https://arxiv.org/abs/1803.03599}.	

\bibitem{ll1} A. Majhi, {\it A Logico-Linguistic Inquiry into the Foundations of Physics: Part 1},  \href{ https://doi.org/10.1007/s10516-021-09593-0}{Axiomathes, 32, 153-198
	(2021)}; \href{https://arxiv.org/abs/2110.03514}{https://arxiv.org/abs/2110.03514}.

\bibitem{lpp} A. Majhi, {\it Logic, Philosophy and Physics: A Critical Commentary on the Dilemma of Categories}, \href{https://doi.org/10.1007/s10516-021-09588-x}{Axiomathes, 32, 1415-1431 (2021)}; \href{https://arxiv.org/abs/2110.11230}{https://arxiv.org/abs/2110.11230}.



\bibitem{wm1} A. S. Wightman, {\it Hilbert's sixth problem: Mathematical treatment of the axioms of physics}''(1976) in Felix E. Browder (ed.). Mathematical Developments Arising from Hilbert Problems. Proceedings of Symposia in Pure Mathematics. Vol. XXVIII. American Mathematical Society. pp. 147–240.

\bibitem{wm2}R. F. Streater, A. S. Wightman, {\it PCT, Spin and Statistics, and All That}, New York: W. A. Benjamin (1964).

\bibitem{aqft}N. Bogoliubov, A. Logunov, I. Todorov, {\it  Introduction to Axiomatic Quantum Field Theory}. Reading, Massachusetts: W. A. Benjamin (1975).

\bibitem{massgap}A. Jaffe, E. Witten, {\it Quantum Yang-Mills theory}, \href{https://www.claymath.org/sites/default/files/yangmills.pdf}{https://www.claymath.org/sites/default/files/yangmills.pdf}.


\bibitem{brouwer}{\color{black} L. E. J. Brouwer, {\it Collected Works, vol.1 - Philosophy and foundations of mathematics}, Elsevier Science Publishing (1975).}

\bibitem{brouwerconscious} L. E. J. Brouwer, {\it Consciousness, philosophy, and mathematics}, Proceedings of the Tenth International Congress of Philosophy (Amsterdam, August 11–18, 1948), North-Holland Publishing Company, Amsterdam1949, pp. 1235–1249.


\bibitem{brouwerintform} L. E. J. Brouwer,  {\it Intuitionism and Formalism}, Bull. Amer. Math. Soc. 20(2): 81-96 (1913).


\bibitem{goedelincom}K. Goedel, {\it On formally undecidable propositions of Principia Mathematica and related systems I} (1931) in  {\it Kurt Goedel, Collected Works, Volume 1, Publications 1929-1936}, edited by  S. Feferman, J. W. Dawson Jr., S. C. Kleene, G. H. Moore, R. M. Solovay, J. Heijenoort, Oxford University Press, New York; Clarendon Press, Oxford (1986).

\bibitem{turing1} A. Turing, {\it On Computable Numbers, with an application to the Enstscheidungsproblem}, Proceedings of the London Mathematical Society, Vol. s2-42, Issue 1, pp. 230-265 (1937); {\it A correction}, Proceedings of the London Mathematical Society, Vol. s2-43, Issue 1, pp. 544-546 (1938).

\bibitem{turing2} A. Turing, {\it Systems of Logic Based on Ordinals}, Proceedings of the London Mathematical Society (1939).	

\bibitem{church1}{\color{black} A. Church, {\it An unsolvable problem of elementary number theory}, American journal of mathematics, vol. 58 (1936), pp. 345–363; \href{https://doi.org/10.2307/2371045}{https://doi.org/10.2307/2371045}.}

\bibitem{church2} {\color{black}A. Church, {\it A note on the Entscheidungsproblem}, Journal of Symbolic Logic, 1 (1936), pp 40–41; \href{https://doi.org/10.2307/2269326}{https://doi.org/10.2307/2269326}.}

\bibitem{udbook} {\color{black}M. Davis (Ed.), {\it The Undecidable - Basic Papers on Undecidable Propositions, Unsolvable Problems and Computable Functions}, Raven Press Hewlett (1965). }
	
\bibitem{britannica}{\color{black} L. M. Schagrin, G. E. Hughes, {\it Formal logic}. Encyclopedia Britannica,  2023. \href{https://www.britannica.com/topic/formal-logic}{https://www.britannica.com/topic/formal-logic}.}	
	
%\bibitem{complexity1}S. Mertens, {\it Computational Complexity for Physicists}, Computing in Science \& Engineering, vol. 4, no. 3, pp. 31-47, (2002), \href{https://arxiv.org/pdf/cond-mat/0012185.pdf}{https://arxiv.org/pdf/cond-mat/0012185.pdf}.	


\bibitem{costa1} A. da Costa, A. Doria, {\it Undecidability and Incompleteness in Classical Mechanics}, Int. J. Theoret. Phys. 30(8), pp. 1041-1073 (1991). \href{https://doi.org/10.1007/BF00671484}{https://doi.org/10.1007/BF00671484}.

\bibitem{costa2} A. da Costa, A. Doria, {\it Classical Physics and Pernrose's Thesis}, Found. of Phys. Letters 4(4), pp. 343-373 (1991). \href{https://doi.org/10.1007/BF00665895}{https://doi.org/10.1007/BF00665895}.

\bibitem{costa3} N. C. A. da Costa, F. A. Doria, {\it Undecidability, incompleteness and Arnol'd problems},  
Studia Logica, Vol. 55, No. 1, pp. 23-32 (1995).

\bibitem{moore1} C. Moore, {\it Unpredictability and Undecidability in Dynamical Systems}, Phys. Rev. Lett. 64, 2354 (1990). \href{https://www2.seas.gwu.edu/~simhaweb/iisc/Moore.pdf}{https://www2.seas.gwu.edu/~simhaweb/iisc/Moore.pdf}.

\bibitem{moore2}{\color{black}C. Moore, {\it Generalized shifts: unpredictability and undecidability in dynamical systems},  Nonlinearity, 4, 199 (1991).\href{https://sites.santafe.edu/~moore/nonlinearity-gs.pdf}{https://sites.santafe.edu/~moore/nonlinearity-gs.pdf}.
}

\bibitem{kanter1990} {\color{black} I. Kanter, {\it Undecidability Principle and the Uncertainty Principle Even for Classical Systems}, Phys. Rev. Lett. 64, 332 (1990). \href{https://doi.org/10.1103/PhysRevLett.64.332}{https://doi.org/10.1103/PhysRevLett.64.332}.}


\bibitem{penrose} R. Penrose, {\it The Emperor's New Mind}, Oxford University
Press, Oxford (1989).
	
%\bibitem{gh}D. Gosset, Y. Huang, {\it Correlation Length versus Gap in Frustration-Free Systems}, \href{https://doi.org/10.1103/PhysRevLett.116.097202}{Phys. Rev. Lett. 116, 097202 (2016)},  \href{https://arxiv.org/abs/1509.06360v3}{https://arxiv.org/abs/1509.06360v3}.

\bibitem{udqua1}M. M. Wolf, T. S. Cubitt, D. Perez-Garcia, {\it Are problems in Quantum Information Theory (un)decidable?},  \href{https://arxiv.org/abs/1111.5425}{https://arxiv.org/abs/1111.5425}.

\bibitem{udqua2}J. Eisert, M. P. Müller, C. Gogolin, {\it Quantum measurement occurrence is undecidable}, \href{https://doi.org/10.1103/PhysRevLett.108.260501}{Phys. Rev. Lett. 108, 260501 (2012)}\href{https://arxiv.org/abs/1111.3965}{https://arxiv.org/abs/1111.3965}.

\bibitem{udgap1}T. S. Cubitt, {\it Frustratingly Undecidable (or Undecidably
Frustrating)}, in Proceedings of IQC Waterloo, (2011).

\bibitem{udgap2}T. S. Cubitt, D. Perez-Garcia, M. M. Wolf, {\it Undecidability of the spectral gap},  \href{https://www.nature.com/articles/nature16059}{Nature, vol.528, pp 207 - 211 (2015)},  \href{https://arxiv.org/abs/1502.04573}{https://arxiv.org/abs/1502.04573}.

\bibitem{udgap3}J. Bausch, T. S. Cubitt, A. Lucia, D.  Perez-Garcia, {\it Undecidability of the Spectral Gap in One Dimension},  \href{https://doi.org/10.1103/PhysRevX.10.031038}{Phys. Rev. X 10, 031038 (2020)}.

\bibitem{udgap4}{\color{black} T. S. Cubitt, {\it A Note on the Second Spectral Gap Incompleteness Theorem}, \href{https://doi.org/10.48550/arXiv.2105.09854}{https://doi.org/10.48550/arXiv.2105.09854}.}


\bibitem{landsman2020}{\color{black}K. Landsman, {\it Indeterminism and Undecidability} (2020),  \href{https://doi.org/10.48550/arXiv.2003.03554}{https://doi.org/10.48550/arXiv.2003.03554}.}

\bibitem{svozil1993}K. Svozil, {\it Randomness and Undecidability in Physics}, World Scientific (1993); \href{https://www.worldscientific.com/worldscibooks/10.1142/1524}{https://www.worldscientific.com/worldscibooks/10.1142/1524}.	

\bibitem{svozil1995}{\color{black}K. Svozil, {\it Undecidability everywhere?}, \href{https://arxiv.org/abs/chao-dyn/9509023v1}{https://arxiv.org/abs/chao-dyn/9509023v1}.}

\bibitem{calude2007}{\color{black}  C. S. Calude, M. A. Stay, {\it From Heisenberg to G¨odel via Chaitin}, Int J Theor Phys 46, 2013–2025 (2007). \href{https://doi.org/10.1007/s10773-006-9296-8}{https://doi.org/10.1007/s10773-006-9296-8}.}


\bibitem{book2021} {\color{black}A. Aguirre, Z. Merali, D. Sloan (Eds.), {\it Undecidability, Uncomputability, and Unpredictability}, Springer Cham (2021). \href{https://doi.org/10.1007/978-3-030-70354-7}{https://doi.org/10.1007/978-3-030-70354-7}.}

\bibitem{calude}{\color{black} C. Calude, D. I. Campbell, K. Svozil, D. Ştefănecu, {\it Strong Determinism vs. Computability}, \href{https://arxiv.org/abs/quant-ph/9412004}{https://arxiv.org/abs/quant-ph/9412004}.}


\bibitem{kaufmann2019}{\color{black}	M. Prokopenko, M. Harré, J. Lizier, F. Boschetti, P. Peppas, S.  Kauffman, {\it Self-referential basis of undecidable dynamics: from The Liar Paradox and The Halting Problem to The Edge of Chaos}, Physics of Life Reviews, Volume 31, December 2019, Pages 134-156,  \href{https://arxiv.org/abs/1711.02456}{https://arxiv.org/abs/1711.02456}.
}


\bibitem{fraenkel} A. A. Fraenkel, Y. Bar Hillel, A. Levy, {\it Foundations of Set Theory - Studies in Logic and The Foundations of Mathematics, Volume 67}, Elsevier (1973).	

\bibitem{tarskitruth} {\color{black}A. Tarski, {\it The Semantic Conception of Truth and the Foundations of Semantics}, 
	Philosophy and Phenomenological Research Vol. 4, No. 3 (Mar., 1944), pp. 341-376 (36 pages), \href{https://doi.org/10.2307/2102968}{https://doi.org/10.2307/2102968}. \href{http://www.ditext.com/tarski/tarski.html}{http://www.ditext.com/tarski/tarski.html}.}

\bibitem{tarskiundef}{\color{black} A. Tarski, {\it The Concept of Truth in Formalized Languages} in 
{\it Logics, Semantics, Metamathematics - Papers from 1923 to 1938}, translated by J. H. Woodger, Second Edition, Hackett (1983).}

\bibitem{tarskidef}{\color{black} A. Tarski, {\it Introduction to logic and to the methodology of the deductive sciences}, Oxford University Press (1994).}

\bibitem{kripketruth} {\color{black}S. Kripke, {\it Outline of a theory of truth}, Seventy-Second Annual Meeting American Philosophical Association, Eastern Division. Vol. 72. Journal of Philosophy. pp. 690–716. \href{https://doi.org/10.2307/2024634}{https://doi.org/10.2307/2024634}.}

\bibitem{stanexistence}{\color{black} M. Nelson, {\it Existence}, The Stanford Encyclopedia of Philosophy (Winter 2022 Edition), Edward N. Zalta \& Uri Nodelman (eds.), \href{https://plato.stanford.edu/archives/win2022/entries/existence/}{https://plato.stanford.edu/archives/win2022/entries/existence/}.}

\bibitem{wigner}{\color{black} E. P. Wigner, {\it Remarks on the Mind-Body Question}. In: Mehra, J. (eds) Philosophical Reflections and Syntheses. The Collected Works of Eugene Paul Wigner, vol B / 6. Springer, Berlin, Heidelberg.\href{https://doi.org/10.1007/978-3-642-78374-6_20}{https://doi.org/10.1007/978-3-642-78374-6\_20}.} 


\bibitem{lyapunov}{\color{black} V.I. Zubov, {\it Methods of A.M. Lyapunov and their application}, Nordhoff (1964).}

\bibitem{earman2007}{\color{black} J. Earman, {\it Aspects of determinism in modern physics}, in Philosophy of physics Part B, 1369-1434 (2007), \href{https://doi.org/10.1016/B978-044451560-5/50017-8}{https://doi.org/10.1016/B978-044451560-5/50017-8}.}

\bibitem{earman1986}{\color{black} J. Earman, {\it A primer on determinism}, Springer Dordrecht (1986). \href{https://sites.pitt.edu/~jearman/Earman_1986PrimerOnDeterminism.pdf}{https://sites.pitt.edu/~jearman/Earman\_1986PrimerOnDeterminism.pdf}.}

\bibitem{standeterminism}{\color{black}Hoefer, Carl, {\it Causal Determinism}, The Stanford Encyclopedia of Philosophy (Spring 2023 Edition), Edward N. Zalta \& Uri Nodelman (eds.), \href{https://plato.stanford.edu/entries/determinism-causal/}{https://plato.stanford.edu/entries/determinism-causal/}.}

\bibitem{popper}{\color{black} K. Popper, {\it Indeterminism in Quantum Physics and in Classical Physics. Part I}, The British Journal for the Philosophy of Science 1, no. 2 (1950): 117–33. http://www.jstor.org/stable/685807.  \href{https://cqi.inf.usi.ch/qic/popper-indet.pdf}{https://cqi.inf.usi.ch/qic/popper-indet.pdf}.}


\bibitem{gisin1}{\color{black} N. Gisin, {\it  Indeterminism in Physics, classical chaos and
	Bohmian mechanics. Are real numbers really real?}. Erkenn. 86, 1469–1481 (2021), \href{https://doi.org/10.1007/s10670-019-00165-8}{https://doi.org/10.1007/s10670-019-00165-8}.}

\bibitem{gisin2}{\color{black}N. Gisin, {\it Indeterminism in physics and intuitionistic mathematics}, Synthese 199, 13345-13371 (2021); \href{https://philpapers.org/archive/GISIIP-2.pdf}{https://philpapers.org/archive/GISIIP-2.pdf}.}


\bibitem{gisinhidden}{\color{black} N. Gisin, {\it Real numbers are the hidden variables of classical mechanics}. Quantum Stud.: Math. Found. 7, 197–201 (2020). \href{https://doi.org/10.1007/s40509-019-00211-8}{https://doi.org/10.1007/s40509-019-00211-8}.}

\bibitem{santogisin1}{\color{black} Del Santo, F. and Gisin, N. (2019). {\it Physics without determinism: Alternative interpretations of classical physics}. Physical Review A, 100(6), 062107.}

\bibitem{santogisin2}{\color{black}F. Del Santo, N. Gisin, {\it The Relativity of Indeterminacy}. Entropy, 23, 1326 (2021). \href{https://doi.org/10.3390/e23101326}{https://doi.org/10.3390/e23101326}. }

\bibitem{santo1}{\color{black} F. Del Santo,  {\it  Indeterminism, causality and information: Has physics ever been deterministic?}. In Aguirre, A., Foster, B., and Z. Merali (Eds.), Undecidability, Uncomputability,
	and Unpredictability, Springer Nature (2021), \href{https://doi.org/10.1007/978-3-030-70354-7_5}{https://doi.org/10.1007/978-3-030-70354-7\_5}. }


\bibitem{yami2020}{\color{black}H. Ben-Yami, {\it  The Structure of Space and Time, and
	Physical Indeterminacy} (2020). \href{https://doi.org/10.48550/arXiv.2005.05121}{https://doi.org/10.48550/arXiv.2005.05121}.}

\bibitem{bell}{\color{black} J. S. Bell, {\it On the Einstein Podolsky Rosen paradox}, 
	Physics Physique Fizika 1, 195 (1964). \href{https://doi.org/10.1103/PhysicsPhysiqueFizika.1.195}{https://doi.org/10.1103/PhysicsPhysiqueFizika.1.195}.}

\bibitem{kochenspecker}{\color{black} S. Kochen, E. P. Specker, {\it The problem of hidden variables in quantum mechanics}. Journal of Mathematics and Mechanics. 17 (1): 59–87 (1967). \href{http://dx.doi.org/10.1512/iumj.1968.17.17004}{http://dx.doi.org/10.1512/iumj.1968.17.17004}.}


\bibitem{landsman2014}{\color{black} E. Cator, K. Landsman, {\it Constraints on determinism: Bell versus Conway–Kochen}. Foundations of Physics. 44 (7): 781–791 (2014). \href{https://arxiv.org/abs/1402.1972}{https://arxiv.org/abs/1402.1972}.}

\bibitem{conkoch2006}{\color{black} J. Conway, S. Kochen, {\it The Free Will Theorem}. Foundations of Physics. 36 (10): 1441 (2006).\href{https://arxiv.org/abs/quant-ph/0604079}{https://arxiv.org/abs/quant-ph/0604079}.}

\bibitem{conkoch2009}{\color{black} J. H. Conway, S. Kochen (2009), {\it The strong free will theorem}. Notices of the AMS. 56 (2): 226–232 (2009). \href{https://doi.org/10.48550/arXiv.0807.3286}{https://doi.org/10.48550/arXiv.0807.3286}.}

\bibitem{kochen2017}{\color{black} S. Kochen, {\it Born's Rule, EPR, and the Free Will Theorem} (2017), \href{https://arxiv.org/abs/1710.00868}{https://arxiv.org/abs/1710.00868}.}


\bibitem{heyred1983}{\color{black} P. Heywood, M. L. G. Redhead, {\it Nonlocality and the Kochen-Specker paradox}. Found Phys 13, 481–499 (1983). \href{https://doi.org/10.1007/BF00729511}{https://doi.org/10.1007/BF00729511}.}


\bibitem{landsmanrandom}{\color{black} K. Landsman, {\it Randomness? What Randomness?}. Found Phys 50, 61–104 (2020). \href{https://doi.org/10.1007/s10701-020-00318-8}{https://doi.org/10.1007/s10701-020-00318-8}}.

\bibitem{cs1}{\color{black}A. Aho, J. Ullmann, {\it Foundations of Computer Science - C Edition}, Free Web Version, \href{http://infolab.stanford.edu/~ullman/focs.html}{http://infolab.stanford.edu/~ullman/focs.html}. 
}

\bibitem{cs2}{\color{black} D. Mandrioli, C. Ghezzi, {\it Theoretical Foundations of Computer Science}, John Wiley \& Sons (1987).}


\bibitem{mills1997}{\color{black}I. M. Mills, {\it The Language of Science}, Metrologia, 34, 101 (1997), \href{https://iopscience.iop.org/article/10.1088/0026-1394/34/1/15}{https://iopscience.iop.org/article/10.1088/0026-1394/34/1/15}.}

\bibitem{guggenheim1942}{\color{black}E. A. Guggenheim M.A. Sc.D. (1942) XLIX. Units and dimensions, The London, Edinburgh, and Dublin Philosophical Magazine and Journal of Science, 33:222, 479-496, \href{https://doi.org/10.1080/14786444208521226}{https://doi.org/10.1080/14786444208521226}.}


\bibitem{boer1995}{\color{black} J. de Boer, {\it On the History of Quantity Calculus and the International System}, Metrologia, 31, 405 (1995), \href{https://iopscience.iop.org/article/10.1088/0026-1394/31/6/001}{https://iopscience.iop.org/article/10.1088/0026-1394/31/6/001}.}

\bibitem{williams2014}{\color{black}J. H. Williams, {\it Defining and Measuring Nature - The Make of All Things}, Morgan \& Claypool Publishers (2014), \href{https://doi.org/10.1088/978-1-627-05279-5}{https://doi.org/10.1088/978-1-627-05279-5}.}

\bibitem{mari2009}{\color{black} L. Mari, {\it On (kinds of) quantities}, Metrologia 46 (2009) L11–L15,  \href{http://dx.doi.org/10.1088/0026-1394/46/3/N01}{http://dx.doi.org/10.1088/0026-1394/46/3/N01}.}

\bibitem{marietal2012}{\color{black}L. Mari et al, {\it Quantity and quantity value}, Metrologia 49 (2012) 756–764, \href{http://dx.doi.org/10.1088/0026-1394/49/6/756}{http://dx.doi.org/10.1088/0026-1394/49/6/756}.}

\bibitem{emerson2004}{\color{black}W. H. Emerson, {\it On the algebra of quantities and their units}, Metrologia 41 L33 (2004), \href{https://iopscience.iop.org/article/10.1088/0026-1394/41/6/L02}{https://iopscience.iop.org/article/10.1088/0026-1394/41/6/L02}.}


\bibitem{ehrlich2007}{\color{black} C. Ehrlich et al, {\it Evolution of philosophy and description of measurement (preliminary rationale for VIM3)}, Accreditation and Quality Assurance, volume 12, pages 201–218  (2007),  \href{https://doi.org/10.1007/s00769-007-0259-4}{https://doi.org/10.1007/s00769-007-0259-4}.}

\bibitem{wolff}{\color{black}J. E. Wolff, {\it The metaphysics of quantities}, Oxford University Press (2020).}

\bibitem{vim2007}{\color{black}International Vocabulary of Metrology – Basic and
	General Concepts and Associated Terms (VIM),
	3rd Edition, \href{https://www.oiml.org/en/files/pdf_v/v002-200-e07.pdf}{https://www.oiml.org/en/files/pdf\_v/v002-200-e07.pdf}	}


\bibitem{vim2021}{\color{black}International Vocabulary of Metrology
	Fourth edition – Committee Draft (VIM4 CD) (2021), \href{https://www.bipm.org/documents/20126/54295284/VIM4_CD_210111c.pdf/a57419b7-790f-2cca-f7c9-25d54d049bf6}{https://www.bipm.org/documents/20126/54295284/VIM4\_CD\_210111c.pdf/a57419b7-790f-2cca-f7c9-25d54d049bf6} [``Please note that the contents of this document
	shall not be quoted in any publication'']}

%\bibitem{cipm}International Committee for Weights and Measures (CIPM), \href{https://www.bipm.org/en/committees/ci/cipm}{https://www.bipm.org/en/committees/ci/cipm}.	

%\bibitem{bipm} {\it BIPM: The International System of Units (SI)}, Brochure, 9th Edition (2019), \href{https://www.bipm.org/documents/20126/41483022/SI-Brochure-9-EN.pdf/2d2b50bf-f2b4-9661-f402-5f9d66e4b507?version=1.9&download=true}{online link}.

%\bibitem{maxwell}J. C. Maxwell, {\it A Treatise on Electricity and Magnetism}. Volume 1, Clarendon Press (1873).



\bibitem{bohr1958} {\color{black} N. Bohr, {\it Causality and Complementarity}, pp.1-7 in {\it Essays 1958-1962 on Atomic Physics and Human Knowledge}, Interscience Publishers (1963).} 


\bibitem{chupqg} A. Majhi, {\it Cauchy's Logico-Linguistic Slip, the Heisenberg Uncertainty and a Semantic Dilemma Concerning ``Quantum Gravity''}, \href{https://doi.org/10.1007/s10773-022-05051-8}{Int J Theor Phys 61, 55 (2022)}; \href{https://hal.archives-ouvertes.fr/hal-03597958}{https://hal.archives-ouvertes.fr/hal-03597958}; \href{https://arxiv.org/abs/2204.00418}{https://arxiv.org/abs/2204.00418}. 


\bibitem{maxwellepr}{\color{black}A. Majhi, {\it Unprovability of First Maxwell's Equation in Light of EPR's Completeness Condition -- A Computational Approach from Logico-linguistic Perspective}, Pramana - Journal of Physics (2023) [to be published], \href{https://hal.science/hal-03682283v2}{https://hal.science/hal-03682283v2}.}


\bibitem{dot}{\color{black} A. Majhi, {\it Resolving the singularity by looking at the dot and demonstrating the undecidability of the continuum hypothesis}, Found. Sci. (2022), \href{https://doi.org/10.1007/s10699-022-09875-9}{https://doi.org/10.1007/s10699-022-09875-9},  \href{https://hal.archives-ouvertes.fr/hal-03528767}{https://hal.archives-ouvertes.fr/hal-03528767}.}   

\bibitem{savage}L. J. Savage, {\it The Foundations of Statistics}, New York, Dover Publications (1972).

\bibitem{decision1} M. Resnik, {\it Choices - An Introduction to Decision Theory}, Univ. Minnesota (1987).

\bibitem{decision2} M. Peterson, {\it An Introduction to Decision Theory}, Cambridge University Press (2009).

\bibitem{amrr1}A. Majhi, R. Radhakrishnan, {\it Problem of identity and quadratic equation}, \href{https://hal.archives-ouvertes.fr/hal-03554501v1}{https://hal.archives-ouvertes.fr/hal-03554501v1}.

\bibitem{amrr2}A. Majhi, R. Radhakrishnan, {\it Inadequacy of Classical Logic in Classical Harmonic Oscillator and the Principle of Superposition}, \href{https://hal.archives-ouvertes.fr/hal-03556334v1}{https://hal.archives-ouvertes.fr/hal-03556334v1}.

\bibitem{buddhamath} A. Majhi, {\it The intuitive root of classical logic, an associated decision problem and the middle way},  \href{https://hal.archives-ouvertes.fr/hal-03587270}{https://hal.archives-ouvertes.fr/hal-03587270}.

\bibitem{millikan}R. Millikan, {\it On the elementary electrical charge and the Avogadro constant}, Phys. Rev. 2, 109 (1913), \href{https://journals.aps.org/pr/abstract/10.1103/PhysRev.2.109}{online link}.

\bibitem{fletcher} H. Fletcher, {\it My Work with Millikan on the Oil-drop Experiment},  Physics Today 35, 6, 43 (1982), \href{https://physicstoday.scitation.org/doi/10.1063/1.2915126}{online link}.


\bibitem{purcell} E. Purcell, {\it  Electricity and magnetism}, Mc Graw-Hill (1985).

\bibitem{schwartz} M. Schwartz, {\it Principles of electrodynamics}, Dover Publications (1987).

\bibitem{rmc} J. Reitz, F. Milford, R. Christy, {\it Foundations of Electromagnetic Theory}, Addison Wesley (1992).

%\bibitem{reason} Such is the essence of reasoning that I find in some classic texts e.g. on page no. 45 of ref.\cite{maxwell} or on page no.24 of ref.\cite{jackson}.

\bibitem{jackson}J. D. Jackson, {\it Classical electrodynamics}, Wiley (1999).

\bibitem{bipm} {\it BIPM: The International System of Units (SI)}, Brochure, 9th Edition (2019), \href{https://www.bipm.org/documents/20126/41483022/SI-Brochure-9-EN.pdf/2d2b50bf-f2b4-9661-f402-5f9d66e4b507?version=1.9&download=true}{online link}.

\bibitem{piaget}{\color{black} J. Piaget, {\it Psychology of Intelligence}, Routledge Classics (1999).}


\bibitem{psychophysics} {\color{black} D. King, B. Viney, W. Woody, W. Douglas, {\it A history of psychology - ideas and context}, Psychology Press (2016). }

\bibitem{machanalysis}{\color{black} E. Mach, {\it The Analysis of Sensations and the Relation of the Physical to the Psychical}, Dover Publications (1959).}

\bibitem{psychtime}{\color{black} A. Haigh, D. Apthorp, L. A. Bizo, {\it The role of Weber’s law in human time perception}. Atten Percept Psychophys 83, 435–447 (2021). \href{https://doi.org/10.3758/s13414-020-02128-6}{https://doi.org/10.3758/s13414-020-02128-6}.}


\bibitem{gisinnature}{\color{black} N. Gisin, {\it Mathematical languages shape our understanding of time in physics}, Nat. Phys. 16, 114–116 (2020), \href{http://doi.org/10.1038/s41567-019-0748-5}{http://doi.org/10.1038/s41567-019-0748-5}.}



%\bibitem{bridges}{\color{black}D. S. Bridges, {\it Can Constructive Mathematics be Applied in Physics?}, Journal of Philosophical Logic 28, 439–453 (1999). \href{https://doi.org/10.1023/A:1004420413391}{https://doi.org/10.1023/A:1004420413391}.}


\bibitem{landau}{\color{black} L. D. Landau, E. M. Lifshitz, {\it Fluid Mechanics}, Second Edition, Pergamon Press (1987).}

\bibitem{batchelor} {\color{black}G. K. Batchelor, {\it An Introduction to Fluid Dynamics}, Cambridge University Press (2000).}

\bibitem{marsden}{\color{black}A. J. Chorin, J. E. Marsden, {\it A Mathematical Introduction to Fluid Mechanics}, Springer (1993).}

\bibitem{griffiths}{\color{black} D. J. Griffiths, {\it Introduction to Electrodynamics}, Fourth Edition, Pearson (2015).}

\bibitem{dedekindnum} {\color{black}R. Dedekind, {\it Essays on the Theory of Numbers: (I) Continuity and Irrational Numbers (II) The Nature and Meaning of Numbers}, Dover Publications (1963).}


\end{thebibliography}
\end{document}